\documentclass[conference]{ieeeconf}

\usepackage[all]{xy}
\usepackage{color}
\usepackage{bm}
\usepackage{graphicx}
\usepackage{color}
\usepackage{wrapfig}
\usepackage{epsf}
\usepackage{times,mathptm}
\usepackage{url}
\usepackage{amssymb, amsmath}

\begin{document}

\title{Polytope of Correct (Linear Programming) Decoding\\ and
Low-Weight Pseudo-Codewords}

\author{
\authorblockN{Michael Chertkov}
\authorblockA{CNLS and Theoretical Divison at LANL\\
\& New Mexico Consortium,\\
Los Alamos, NM 87545, USA\\
Email: chertkov@lanl.gov}
\and
\authorblockN{Mikhail Stepanov}
\authorblockA{Department of Mathematics\\
University of Arizona\\
Tucson, AZ 85721, USA\\
Email: stepanov@math.arizona.edu}
}

\maketitle

 \begin{abstract}
We analyze Linear Programming (LP) decoding of graphical binary codes
operating over soft-output, symmetric and log-concave channels. We show
that the error-surface, separating domain of the correct decoding from
domain of the erroneous decoding, is a polytope. We formulate the
problem of finding the lowest-weight pseudo-codeword as a non-convex
optimization (maximization of a convex function) over a polytope, with the cost
function defined by the channel and the polytope defined by the structure of
the code. This formulation suggests new provably convergent heuristics for finding the
lowest weight pseudo-codewords improving in quality upon previously discussed. 
The algorithm performance is tested on the example of the Tanner $[155, 64, 20]$ code over the Additive
White Gaussian Noise (AWGN) channel.
 \end{abstract}

\section{Introduction}
\label{sec:intro}

Low-Density Parity Check (LDPC) codes are capacity achieving (in the thermodynamic limit) and are easy to decode via message-passing algorithms of the Belief-Propagation (BP) type
\cite{61Gal,richardsonurbanke,RiU05}. However, performance of the efficient
decoder on a given finite code is not ideal, resulting in a sizable
difference between optimal (Maximum A-Posteriori) and the suboptimal
decoders observed in the asymptotics of Bit-Error-Rates (BERs) at the
high Signal-to-Noise-Ratios (SNR), in the {\it error floor} regime \cite{03Rich}.
Errors in this extreme regime of the {\it error-floor} are mainly due to
special configurations of the channel noise, called instantons
\cite{Stepanov2005}, correspondent to decoding into pseudo-codewords
\cite{wiberg,01Forney} different from any of the codewords of the code. Analysis of
the instantons and pseudo-codewords in the case of LP decoder \cite{05FWK} is of a special interest. LP is a
combinatorial (zero-temperature) version of BP, thus admitting
convenient description in terms of the pseudo-codeword polytope \cite{05FWK}.  The
geometric structure associated with the polytope gave rise to new decoding techniques related to graph covers \cite{03KV},
adaptive processing of the polytope constraints \cite{08TS}, and the concept of
LP duality \cite{06VK}. The succinct combinatorial formulation of the
coding was also useful in terms of improving LP and thus reducing the
gap between the LP and MAP decoders \cite{09DGW,06CC,07C,07YWF, 09YDW}.

In \cite{08CS} we suggested an LP-specific heuristic Pseudo-Codeword 
Search (PCS) algorithm. The main idea of the algorithm was based on 
exploring the Wiberg relation, from \cite{wiberg,01Forney}, between 
pseudo-codeword and an optimal noise configuration which lies on the 
median between the pseudo-codeword and zero-codeword. In essence, the 
algorithm of \cite{08CS} performs a biased walk over the exterior of the 
domain of correct LP decoding (surrounding zero codeword) and arrives at 
the error-surface (boundary of the domain) in a small finite number of 
steps.  The algorithm, tested on some number of codes over the AWGN 
channel, showed excellent performance. For any noise initiation it 
always approaches the error-surface monotonically in simulations, even 
though the monotonicity proof was not provided. Latter the algorithm was 
generalized to the case of discrete-output channel (specifically Binary 
Symmetric (BS) channel) in \cite{09CCSV,09CVSC}, where the monotonicity 
proof was given. The technique was also extended to discover the most 
probable configurations of error-vectors in compressed sensing 
\cite{10CCV}.

This paper continues the trend of \cite{08CS} and analyzes the
error-surface and the associated low-weight pseudo-codewords. We study
the domain of correct decoding, bounded by the error-surface; formulate
the (channel specific) problem of finding the most probable
configuration of the noise leading to a failure (and respective
pseudo-codeword) as an optimization problem; design an efficient
heuristic; and illustrate performance of
the algorithm on the exemplary Tanner $[155, 64, 20]$ code
\cite{tannercode}. The main statements of the manuscript are:
 \begin{itemize}
  \item The domain of correct decoding is a polytope in the noise space. 
For a typical code the polytope is likely to be non-tractable, {\it 
i.e.}, requiring description exponential in the code size. [Section 
\ref{sec:Polytope}.]
  \item The problem of finding the lowest weight pseudo-codeword of a
graphical code over log-concave symmetric (for example AWGN) channel is
reduced to maximization of a convex function, associated with the
channel, over a polytope, associated with the code and defined as
the cross-section of the decoding polytope by a plane. [Section
\ref{sec:optimization}.]

\item We suggested Majorization Optimization Algorithm (MOA), based on 
majorization-minimization \cite{04HL} approximation of the 
aforementioned optimization formulation. We showed that MOA, as well as 
previously introduced PCS, are both monotonic in discovering iteratively 
the low-weight pseudo-codewords (effective weight decreases with 
iterations). [Section \ref{sec:Cone}.] Performances of MOA and PCS are 
tested on the Tanner code over AWGN channel in Section \ref{sec:tests}.

\end{itemize}

\section{Preliminary Discussions and Definitions}
\label{sec:Preliminary}

We consider LP decoding \cite{05FWK} of  binary LDPC code and discuss the problem of finding
the most probable configuration of the noise, so-called instanton, for
which the decoding fails \cite{08CS}. Equivalently stated, this is the
problem of finding the lowest weight (closest to the zero codeword)
pseudo-codeword of the code.

The technique we discuss here applies to any soft-output, symmetric 
channels where the transition probability, ${\cal P}({\bm 
x}|{\bm\sigma})$,  from the codeword ${\bm \sigma}$ to the channel 
output ${\bm x}$, is a log-convex function of ${\bm x}$, {\it i.e.}, 
$-\log({\cal P}({\bm x}|{\bm\sigma}))$ is a convex function of ${\bm x}$). AWGN channel is our enabling example with
 \begin{eqnarray}
  {\cal P}({\bm x}|{\bm\sigma})\propto
\exp\left(-2s^2\sum_{i=1}^N(x_i-\sigma_i)^2\right),
  \label{AWGN}
 \end{eqnarray} where $s$ is the signal-to-noise ratio of the noise,
${\bm \sigma}=(\sigma_i=0,1|i=1,\cdots, N)$, is the binary $N$-bits long
codeword launched into the channel, and ${\bm x} = (x_i \in {\bf
R}|i=1,\cdots, N)$ is the real valued signal received by the decoder.

Maximum Likelihood decoding can be formulated as an LP optimization over
the polytope, ${\cal P}$, spanned by all the codewords of the code ${\it
C}$,
 \begin{eqnarray}
  \min_{{\bm \sigma'}} \sum_i(1-2 x_i)\sigma_i' \bigg|_{{\bm \sigma'}\in
{\cal P}} \ .
  \label{ML_LP}
 \end{eqnarray} 
However, the full codeword polytope is exponentially
large in the code size and thus it is not tractable. Trading optimality
for efficiency the authors of \cite{05FWK} have suggested to relax the
full polytope into a tractable one (stated in terms of a polynomial, in
the size of the code, number of constraints).  The relaxation, coined
LP-decoding, is based on decomposition of the code into small individual
checks based codes thus assuring (by construction) that the set of
original codewords forms a subset of all the corners of the relaxed
polytope (so-called set of pseudo-codewords). The LP-decoding can be
formulated in multiple ways. Following \cite{08CS}, we choose to start here with the formulation of LP,
correspondent to the so-called zero-temperature version of the Bethe
Free Energy approach of \cite{05YFW}:
 \begin{eqnarray}
  && LP_p({\bm x}) = \min_{{\bm b}}\sum_i(1-2
x_i)\sum_{\sigma_i=0,1}\sigma_i b_i(\sigma_i)\bigg|_{ {\bm b}\in {\cal
P}_l},
  \label{BP_LP}\\
  && {\cal P}_l = \left\{\begin{array}{c} \forall i:\
\sum_{\sigma_i}b_i(\sigma_i)=1;\ \forall \alpha:\ \sum_{{\bm
\sigma}_\alpha}b_\alpha({\bm\sigma}_\alpha)=1; \\ \forall i,\ \forall
\alpha\sim i:\ \sum_{\sigma_i}(1-2\sigma_i)b_i(\sigma_i)\\= \sum_{{\bm
\sigma}_\alpha}(1-2\sigma_i)b_\alpha({\bm\sigma}_\alpha);\\ \forall i:\
b_i(\sigma_i)\geq 0;\ \forall \alpha:\ b_\alpha({\bm\sigma}_\alpha)\geq
0\end{array}\right\}, \nonumber
 \end{eqnarray}
where $b$ are beliefs, {\it i.e.}, proxies for respective
marginal probabilities. ${\cal P}_l$ is a polytope, which we call large
(LP-decoding) polytope. ${\cal P}_l$ only depends on the structure (graph) of
the code (and it does not depend on the channel model). There are
beliefs of two types associated with two types of nodes in the parity
check graph of the code, ${\cal G}$, bits $i$ and checks $\alpha$
respectively. $\sigma_i=0,1$ represent values of the bit $i$, and the
vector ${\bm\sigma}_\alpha=(\sigma_i|i\sim\alpha; \mbox{s.t. }
\sum_i\sigma_i=0\mbox{ mod }2)$ stands for one of the allowed local
codewords associated with the check $\alpha$. Of the conditions in the
definition of ${\cal P}_l$, the first two equalities are normalizations
(for the beliefs/probabilities), the third equality states
consistency between beliefs associated with bits and checks. The two
last inequalities in ${\cal P}_l$ ensure that the beliefs
(probabilities) are positive. If the channel noise corrupting the
zero codeword is sufficiently weak, {\it i.e.}, if $|{\bm x}|\ll 1$, the 
$LP_p$
outputs zero, corresponding to successful decoding. However, $LP_p$
confuses another pseudo-codeword (typically non-integer) for the
codeword if ${\bm x}$ is sufficiently noisy, then giving a strictly
negative output, $LP_p<0$.

Description of the $LP_p({\bm x})$ in Eq.~(\ref{BP_LP}) can be restated
in terms of a smaller set of beliefs, only bit beliefs $\pmb{\beta} =
(\beta_i=b_i(1)|i=1,\cdots,N)$. Then the ``small polytope" formulation
of Eq.~(\ref{BP_LP}) becomes \cite{Yannakakis91,05FWK}:
 \begin{eqnarray}
  && LP_p({\bm x}) = \min_{\pmb{\beta}} \sum_i (1 - 2 x_i) \beta_i
\Big|_{\pmb{\beta} \in {\cal P}_s}, \label{BP_LP1}\\
  && \hspace{-0.3cm}{\cal P}_s \!=\! \left\{\!\! \begin{array}{c} \displaystyle \forall \alpha
\, \forall I \subseteq I_\alpha,  |I|\mbox{ is odd}:\  \sum_{i
\in I} \beta_i \!-\! \sum_{i \in I_\alpha \backslash I} \beta_i \le |I|\! -\! 1 \\
\forall i:\ 0 \le \beta_i \le 1 \end{array}\!\!\right\}, \nonumber
 \end{eqnarray}
where $I_\alpha$ is the subset of bit-nodes contributing check $\alpha$.

The ``large polytope" formulation of the LP-decoding (\ref{BP_LP}) can
also be restated in terms of its dual (the formulation here is almost
identical to DLPD2 of \cite{06VK})
 \begin{eqnarray}
  && LP_d({\bm x})=\left.\max_{{\bm \theta, \phi,\lambda}} \sum_i
\phi_i+\sum_\alpha \theta_\alpha\right|_{ {\bm \theta,
\phi,\lambda}\in{\cal P}_d},
  \label{LP_dual}\\
  && \hspace{-0.3cm}{\cal P}_d\!=\!\left\{\begin{array}{c}\forall i,\ \forall
\sigma_i:\ \sigma_i\!(1\!-\!2x_i)\!-\!(1\!-\!2\sigma_i)\sum_{\alpha\sim
i}\!\lambda_{i\alpha}\geq\phi_i\\ \forall \alpha,\ \forall
{\bm\sigma}_\alpha:\
\sum_{i\sim\alpha}\lambda_{i\alpha}(1-2\sigma_i)\geq
\theta_\alpha\end{array}\right\}, \nonumber
 \end{eqnarray}
where ${\bm \phi}=(\phi_i|i=1,\cdots,N)$, ${\bm
\theta}=(\theta_\alpha|\alpha=1,\cdots,M)$,
${\bm\lambda}=(\lambda_{i\alpha}|(i,\alpha)\in{\cal G}_1)$ are
Lagrangian multipliers (messages) conjugated to the first, second and
third conditions in the original LP (\ref{BP_LP}) respectively.
According to the main (strong duality) theorem of the convex
optimization (see many textbooks, {\it e.g.}, \cite{04BV}) the results 
of 
the
primal problem (\ref{BP_LP}) and the dual problem (\ref{LP_dual})
coincide, $LP_p = LP_d$. 

In this manuscript we are mainly concerned with the following practical
problem: given a finite code, log-concave channel (for concreteness and
without loss of generality we will consider AWGN channel as an example),
and the LP-decoding (in its primal or dual versions), to find the most
probable configuration (instanton) of the channel noise, ${\bm x}$,
imposed on the zero codeword, ${\bm
\sigma}_0={\bm 0}$, which leads to incorrect decoding. Formally, we are solving the following ``instanton" problem
 \begin{eqnarray}
 \min_{\bm x} \sum_i x_i^2 \bigg|_{{\bm x} \in {\cal D}_{ext}},
  \label{inst1}
 \end{eqnarray} where ${\cal D}_{ext}$ is defined as an exterior
(complement) of the domain, ${\cal D}_{int}$, correspondent to the
correct decoding: $LP_p=LP_d=0$. Thus, ${\cal D}_{ext} = {\bf
R}^N\setminus{\cal D}_{int}$.

\section{Domain of Correct Decoding is a Polytope}
\label{sec:Polytope}

Let us show that {\it ${\cal D}_{int}$ is actually a polytope}.

Consider the following auxiliary domain of $({\bm x};{\bm \theta,
\phi,\lambda})$:
 \begin{eqnarray}
  {\cal F}_d = \left\{ \begin{array}{l} \sum_i \phi_i+\sum_\alpha
\theta_\alpha = 0 \\ \forall i,\ \forall \sigma_i:\
\sigma_i(1-2x_i)-(1-2\sigma_i)\sum_{\alpha\sim i}\lambda_{i\alpha}\geq
\phi_i\\ \forall \alpha,\ \forall {\bm\sigma}_\alpha:\
\sum_{i\sim\alpha}\lambda_{i\alpha}(1-2\sigma_i)\geq \theta_\alpha
\end{array}\right\}, \nonumber 
 \end{eqnarray}
constructed from the feasibility region of the dual
problem, $LP_d$, with the zero cost function constraint added.
For any ${\bm x}\in {\cal D}_{int}$ there obviously exists an extended
configuration $({\bm x},{\bm \theta}, {\bm \phi},{\bm \lambda})$ from
${\cal F}_d$. On the other hand, if ${\bm x}\in {\cal D}_{ext}$, then
$LP_p = LP_d < 0$ ({\it i.e.}, a pseudo-codeword, different from the 
zero
codeword, is selected by the LP), and since $LP_d$ is defined as a
maximum over an extension of ${\cal F}_d$ (where the first condition in
${\cal F}_d$ is removed) there exists no valid $({\bm x},{\bm \theta},
{\bm \phi},{\bm \lambda})$ from ${\cal F}_d$ in this case. One concludes
that ${\cal D}_{int}({\bm x})$ coincides with the projection of ${\cal
F}_d$ on the ${\bm x}$ variable
 \begin{eqnarray}
  {\cal D}_{int} = \mbox{Proj} \left( {\cal F}_d \right)_{\bm x} = \{
\exists({\bm \theta, \phi, \lambda}) \mbox{ s.t. } ({\bm x};{\bm \theta,
\phi,\lambda}) \in {\cal F}_d \} .
  \label{Proj}
 \end{eqnarray}
However both ${\cal F}_d$ and its projection to $x$ are
polytopes, {\it i.e.}, ${\cal D}_{int}$ is also a convex domain, 
moreover 
it is
a polytope \footnote{We are thankful to P.~Vontobel for pointing out, after reading the first version of the manuscript,  that the statement above is closely related to these made in \cite{05VK}. See Fig. 11,12 of \cite{05VK} as well as preceding and following discussions.}.

Note that the projected polytope is most likely non-tractable, in the
sense that the number of constraints required to describe the polytope is
expected to be exponential in the dimension of ${\bm x}$ (size of the
code).

\section{Search for Lowest Weight Pseudo-Codeword as an Optimization}
\label{sec:optimization}

Noticing, that Eq.~(\ref{inst1}) is stated in terms of the exterior
domain, ${\cal D}_{ext}$, which is a compliment of ${\cal D}_{int}$, one
attempts to formulate a closely related 
problem stated in terms of optimization over a convex sub-domain of
${\cal D}_{int}$:
 \begin{eqnarray}
  Q(\epsilon)=\left.\min_{\bm x} LP({\bm x})\right|_{{\bm x}\in
\mbox{Ball}_\epsilon},
  \label{min-of-LP}
 \end{eqnarray} where $\mbox{Ball}_\epsilon\equiv\{ \pmb{\zeta} \in {\bf
R}^N:\ \|\pmb{\zeta}\|_2\leq \epsilon\}$ is the ball of radius
$\epsilon$ (which is convex by construction). For sufficiently small
$\epsilon$ any $LP({\bm x})=0$ for any ${\bm x}\in
\mbox{Ball}_\epsilon$, while a gradual increase in $\epsilon$ will
eventually lead, at some $\epsilon_*$, to appearance of the closest to
the zero codeword (in terms of the $l_2$ norm of the AWGN channel) noise
configuration, ${\bm x}_{inst}$, for which $LP({\bm x}_{inst})\leq 0$.
One concludes that the
function of a single parameter, $Q(\epsilon)$, jumps from zero at
$\epsilon<\epsilon_*$ to some negative value at $\epsilon=\epsilon_*$.
Then, $4\epsilon_*^2$ becomes the effective distance of the code (under
the LP-decoding), and the optimal value, ${\bm x}_*$ of $Q(\epsilon_*)$,
corresponds to the most probable instanton.

Using primal formulation of LP-decoding from Eq.~(\ref{BP_LP1}) and
combining minimization over ${\bm x}$ and $\pmb{\beta}$ variables, one
reformulates Eq.~(\ref{min-of-LP}) as the following optimization problem
 \begin{eqnarray}
  && Q(\epsilon)=\left.\min_{\pmb{\beta},{\bm x}} \sum_i(1-2
x_i)\beta_i\right|_{ {\bm x}\in \mbox{Ball}_\epsilon,\
\pmb{\beta}\in{\cal P}_s}.
  \label{min-interior}
 \end{eqnarray}
One important advantage of this formulation is in the fact that
Eq.~(\ref{min-interior}) is stated as an optimization problem, in contrast with the sequential instanton search optimization of \cite{08CS}, where one optimizes over the
noise, then evaluates an internal minimization (the LP decoding itself)
for each configuration of the noise. Note that the cost function in
Eq.~(\ref{min-interior}) is quadratic and concave.

\begin{figure*}[t]
 \centering
 \includegraphics[width=\textwidth]{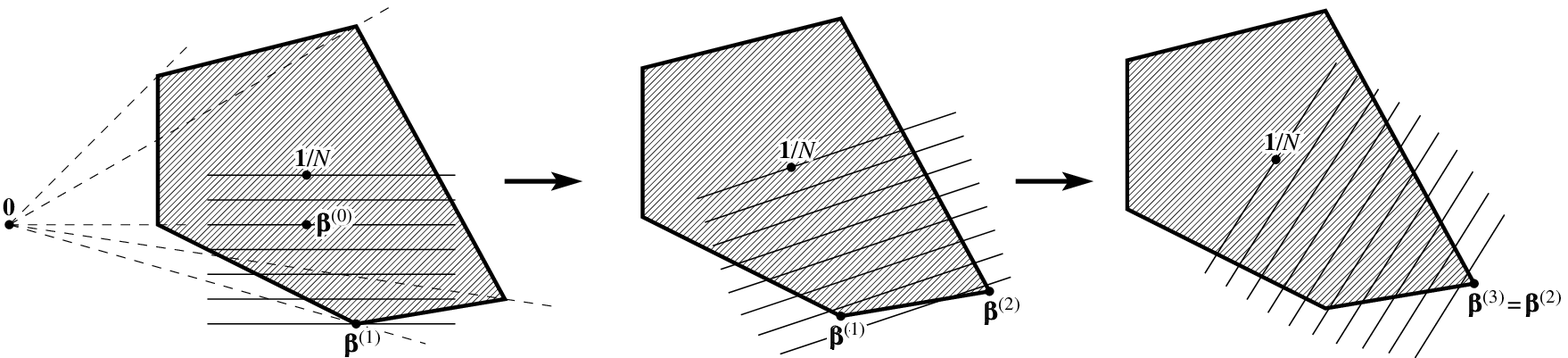}
 \caption{The Figure illustrates the sequential progress (from left to right) of the majorization-minimization procedure. The shaded area corresponds to ${\cal P}_{\rm cone}$. Dashed lines on the left sub-figure show
edges of ${\cal P}_s$ containing the origin ${\bm 0}$.
The optimization starts from $\pmb{\beta}^{(0)}$. Thin solid lines are the
level curves of the linear function $L \big( \pmb{\beta}; \,
\pmb{\beta}^{(k)} \big)$, which optimum  over $\pmb \beta$ results in
$\pmb{\beta}^{(k + 1)}$. The procedure continues till convergence,
$\pmb{\beta}^{(k + 1)} = \pmb{\beta}^{(k)}$ (achieved with $k=2$ in the illustration).}
 \label{fig:moa}
\end{figure*}

Eq.~(\ref{min-interior}) can be simplified further. We expect that the
extremal value  will be achieved (at least for sufficiently large 
$\epsilon$) ``at the surface'' of the ball, {\it i.e.}, at $\sum_i 
x_i^2=\epsilon^2$.  Replacing ${\bm x}\in \mbox{Ball}_\epsilon$ by this
equality and performing optimization over 
${\bm x}$, we arrive (with the help of the standard Lagrangian multiplier
technique, and also assuming that all components of the candidate noise vector are positive)
at the following nonlinear optimization problem stated primarily in terms of the beliefs
 \begin{eqnarray}
  Q(\epsilon)=\left.\min_{\pmb{\beta}} \left(\sum_i \beta_i-2\epsilon
\sqrt{\sum_i \beta_i^2}\right) \right|_{\pmb{\beta} \in {\cal P}_s}.
  \label{min-nonlinear}
 \end{eqnarray}
This problem can be solved approximately (but
efficiently) via the majorization-minimization iterative method
\cite{04HL}, consisting in upper-bounding the cost function by its
linearized expression, minimizing the upper-bound, and iterating by
shifting the linearization point to the solution received on the previous
step. The linearization (for majorization) at each iterative step is
justified because of the following obvious inequality
 \begin{eqnarray}
  && \Vert \pmb{\beta} \Vert_2 \geq L \big( \pmb{\beta}; \,
\pmb{\beta}^{(k)}) = \big( \pmb{\beta} \cdot \pmb{\beta}^{(k)} \big) /
\Vert \pmb{\beta}^{(k)} \Vert_2, \label{major}
 \end{eqnarray} which holds for any $\pmb{\beta}$. Then the iterative
solution of Eq.~(\ref{min-nonlinear}) becomes
 \begin{eqnarray}
   Q^{(k + 1)}(\epsilon) = \min_{\pmb{\beta}} \bigg( \sum_i \beta_i - 2
\epsilon L \big( \pmb{\beta}; \, \pmb{\beta}^{(k)} \big) \bigg)
\bigg|_{\pmb{\beta} \in {\cal P}_s},
  \label{min-lin-iterat}
 \end{eqnarray} where $k = 0,1,\cdots$ till convergence,
$\pmb{\beta}^{(k)}$ is the optimal solution of the optimization found at
the $k$ iteration step, and $\pmb{\beta}^{(k + 1)}$ becomes the optimal
solution at the $(k + 1)$-th iteration. The optimization problem on the
rhs of Eq.~(\ref{min-lin-iterat}) is an LP, {\it i.e.}, it can be solved
efficiently. We also expect that the iterations over $k$ converge fast.
The iterative procedure will depend on the initiation set at
$k = 0$, and starting from different initial conditions we sample
different local optima.

Note that $\epsilon_*$, defined as the smallest $\epsilon$ for
which $Q(\varepsilon)$ becomes negative, allows useful interpretation in
terms of the effective distance of the corresponding pseudo-codeword.
Indeed, $4\varepsilon_*^2 = w(\pmb{\beta})$, where
 \begin{eqnarray}
   w(\pmb{\beta})=\frac{(\sum_i \beta_i)^2}{\sum_i \beta_i^2}.
 \label{weight}
 \end{eqnarray} is the weight of the noise (and of the corresponding
pseudo-codeword) according to the Wiberg formula from
\cite{wiberg,01Forney}, expressing relation between the
direction of the optimal noise and the distance along the direction from
the zero codeword to the error-surface.

\section{Cone Formulation and Majorization Optimization Algorithm}
\label{sec:Cone}

To utilize Eq.~(\ref{min-nonlinear}) for finding the low-weight
pseudo-codewords one needs to scan over the values of $\epsilon$, thus
making one-parametric optimization (over $\epsilon$) in addition to the
(multi-dimensional) optimization contained in Eq.~(\ref{min-nonlinear}).
The main result of this Subsection is that this additional degree
of freedom in the optimization is unnecessary, thus leading to a
simplification of Eq.~(\ref{min-nonlinear}).

Let us first show that: {\it the vertex of ${\cal P}_s$, correspondent to the
pseudo-codeword with the lowest weight, is connected by an edge to the
vertex correspondent to the zero-codeword}.

Since the weight-function, $w(\pmb{\beta})$ from Eq.~(\ref{weight}),
does not depend on the length of the vector $\pmb{\beta}$, one considers
$\pmb{\beta}$, as the direction in the respective space pointing from
the origin, ${\bm 0} = (0,\cdots,0)$, to a point within the polytope
${\cal P}_s$. It is convenient to parameterize the direction in terms of
the projection to the $\sum_i\beta_i=1$ plane. Pseudo-codewords correspond
to special values of the $\pmb{\beta}$ vector projected to the plane,
and to find the pseudo-codeword with the minimum weight we will
need to minimize the weight, $w(\pmb{\beta}) = 1
/\sum_i \beta_i^2$, over the cross-section of the polytope by the
plane (projection). One restates the problem as maximization of $\sum_i
\beta_i^2$, which is also equivalent to finding $\pmb{\beta}$ maximizing
the distance to the central point of the plane within the polytope ${\cal P}_s$, ${\bm 1} / N = (1, \cdots, 1) / N$. ${\cal P}_s$ is projected through the origin to the plane, thus
forming a polytope too (call it cone polytope)
 \begin{eqnarray}
  && {\cal P}_{\rm cone} = \left\{ \begin{array}{c}
 \forall i:\ \beta_i \ge 0 \\
 \displaystyle \forall \alpha \, \forall i \sim \alpha:\ \beta_i \le
\sum_{j \sim \alpha. j \ne i} \beta_j \\
 \sum_i \beta_i = 1
\end{array}\right\}. \label{cone}
 \end{eqnarray} (The projection is understood in the standard projective
space sense, with a line connecting a point within the polytope with the
point of origin, $(0,\cdots,0)$, projecting to the point where the line
crosses the plane.) Note that only faces of  ${\cal
P}_s$ in Eq.~(\ref{BP_LP1}) with $|I| = 1$ become faces of the cone
polytope, ${\cal P}_{\rm cone}$. Further, maximum of $\sum_i \beta_i^2$
is attained at some vertex of the polytope. By construction this vertex
corresponds to an edge connecting the point of origin, $(0,\cdots,0)$
with another vertex of the original polytope ${\cal P}_s$, correspondent
to a pseudo-codeword with the lowest weight. All the other vertexes of
${\cal P}_s$, which are not connected to the origin, are projected to 
interior points of the cone polytope ${\cal P}_{\rm cone}$, thus showing
a higher value of the weight.

The choice of the cone cross-section in Eq.~(\ref{cone}) is convenient
for the purpose of simplifying the optimization problem
(\ref{min-nonlinear}). It guarantees that the first term in the
objective of Eq.~(\ref{min-nonlinear}) is constant, and thus the term is
inessential for the purpose of optimization. In the result, we arrive at
the following reduced version of Eq.~(\ref{min-nonlinear}) (one less
degree of freedom and simpler polytope)
 \begin{eqnarray}
 \tilde{Q} = \max_{\pmb{\beta}} \sqrt{\sum_i \beta_i^2}
\bigg|_{\pmb{\beta} \in {\cal P}_{\rm cone}}.
  \label{cone-opt}
 \end{eqnarray}
According to the discussion above, solution of
Eq.~(\ref{cone-opt}) only describes the optimal direction in the noise
space, ${\bm x}$, and the respective length is reconstructed from the
weight relation $4\epsilon_*^2 = w(\pmb{\beta})$. Thus our final
expression for the optimal noise (instanton), correspondent to the
(optimal) solution of Eq.~(\ref{cone-opt}) is
 \begin{eqnarray}
  {\bm x}= \pmb{\beta} \frac{\sum_i\beta_i}{2\sum_i\beta_i^2}.
 \label{normalization}
 \end{eqnarray}
The geometrical essence of the cone construction and of
the majorization-minimization procedure is illustrated in
Fig.~\ref{fig:moa}.

Few remarks are in order. First, note that there is some additional
freedom in choosing the objective function in the optimization over
$\pmb{\beta}$. For example, one can replace, $\sqrt{\sum_i \beta_i^2}$,
under the sum in Eq.~(\ref{cone-opt}) by $\sum_i (\beta_i - \sum_j
\beta_j / N)^2$, and the resulting optimal $\pmb{\beta}$ stays the same.
Second, the majorization-minimization procedure of
Eq.~(\ref{min-lin-iterat}) for Eq.~(\ref{min-nonlinear}), extends
straightforwardly to any appropriate choice of the objective function in
the reduced optimization, in particular the choice of
Eq.~(\ref{cone-opt}), thus resulting in the sequence 
 \begin{eqnarray}
 \pmb{\beta}^{(k + 1)} = \operatorname*{arg\,max}_{\pmb{\beta}} \,
\pmb{\beta} \cdot \pmb{\beta}^{(k)} \bigg|_{\pmb{\beta} \in {\cal
P}_{\rm cone}} .
  \label{cone-iterat}
 \end{eqnarray}
Third, the sequence (\ref{cone-iterat}) is monotonic by
construction, {\it i.e.}, the effective distance can only decrease with 
the
iteration number $k$, thus proving convergence.

The considerations above suggest the following {\it
Majorization-Optimization Algorithm} (MOA): \begin{itemize}
 \item {\bf Start:} Initiate a point $\pmb{\beta}^{(0)}$ inside the cone
cross-section ${\cal P}_{\rm cone}$ with a random deviation from the
$(1, 1, ..., 1) / N$. [The sampling step.]

 \item {\bf Step 1:} Construct a linear function with the gradient
vector pointing from $(1, 1, ..., 1) / N$ to $\pmb{\beta}^{(k)}$,
optimize it inside ${\cal P}_{\rm cone}$ according to
Eq.~(\ref{cone-iterat}), and get the new $\pmb{\beta}^{(k + 1)}$. [The
majorization-minimization step.]

\item {\bf Step 2:} If $\pmb{\beta}^{(k + 1)} \ne \pmb{\beta}^{(k)}$,
then go to {\bf Step 1}.

\item {\bf End:} Output the optimal noise configuration according to
Eq.~(\ref{normalization}). \end{itemize}

Like PCS of \cite{08CS}, MOA is sensitive to the choice of the initial 
direction in the $\pmb{\beta}$ space, and this clarifies importance of 
repeating sampling step multiple times. Obviously, an individual 
sampling event outputs only pseudo-codewords sharing an edge in ${\cal 
P}_s$ with the zero-codeword, call them ``nearest-neighbors", thus 
ignoring other pseudo-codewords, for example these which are 
``next-nearest-neighbors" to the zero codeword, {\it i.e.}, ones sharing 
an 
edge with a pseudo-codeword which shares an edge with the zero-codeword. 
Even though the effective distance of these ``next-nearest-neighbors" 
may be smaller than the effective distance of some of the 
``nearest-neighbors", MOA guarantees that the exact solution of 
Eq.~(\ref{cone-opt}) can only be a ``nearest-neighbor".

In the remainder of the Section let us briefly compare MOA with PCS. The 
iterative procedure of PCS is analogous to Eq.~(\ref{cone-iterat}) and 
it can be restated as
 \begin{eqnarray}
   \pmb{\beta}^{(k + 1)} = \operatorname*{arg\,max}_{\pmb{\beta}} \,
    \pmb{\beta} \cdot \underbrace{\left( \pmb{\beta}^{(k)}
    \frac{\sum_{i} \beta^{(k)}_i} {\sum_i (\beta^{(k)}_i)^2} - {\bm 1}
    \right)}_{-{\bm h} = 2 {\bm x} - {\bm 1}} \bigg|_{\pmb{\beta} \in
    {\cal P}_s} .
   \label{iterative_PCS}
 \end{eqnarray} 
Note, that, $
   \partial w(\pmb{\beta})/\partial \pmb{\beta}
   = (2\sum_{i} \beta_i)(\sum_i \beta_i^2) \cdot {\bm h}$, so clearly, 
PCS aims to approximate $w(\pmb{\beta})$, linearly inside the polytope, 
${\cal P}_s$.
The function $w(\pmb{\beta})$ is a homogeneous function of degree $0$.
MOA takes advantage of this fact and attempts 
to minimize $w(\pmb{\beta})$ in the projective space of $\pmb{\beta}$,
indexed by the points of ${\cal P}_{\rm cone}$. The value of 
$\max$ in
(\ref{iterative_PCS}) is non-negative, and it is exactly zero at 
$\pmb{\beta} = \pmb{\beta}^{(k)}$. If $w(\pmb{\beta}) <
N$, vector $\partial w(\pmb{\beta}) / \partial \pmb{\beta}$
points away from the central direction ${\bm 1}$, and thus minimization
(\ref{iterative_PCS}) is not going to increase $w(\pmb{\beta})$, 
{\it i.e.}, 
under this (weak and easy to realize) 
condition the PCS is provably monotonic. Also, as $\pmb{\beta} \cdot 
(\partial w(\pmb{\beta}) /
\partial \pmb{\beta}) = 0$, PCS, like MOA, always converges to vertices 
of ${\cal
P}_s$ which are the ``nearest-neighbors" of the zero-codeword (the cone origin).

Since PCS works with $\partial w(\pmb{\beta}) / \partial
\pmb{\beta}$, and not directly with $w(\pmb{\beta})$ like MOA, it ``confuses" $w(\pmb{\beta})$ for
being a homogeneous function of degree $1$. Therefore, compared to MOA, 
PCS has an
additional bias away from the cone origin, thus suggesting that its 
convergence is slower and resulting end-points being further away from 
the cone origin. This assessment is confirmed in the simulations of the 
next Section (see, {\it e.g.}, Fig.~\ref{fig:spectra}).

\section{Tanner Code Test}
\label{sec:tests}

\begin{figure}[t]
\centering
 \includegraphics[width=\columnwidth]{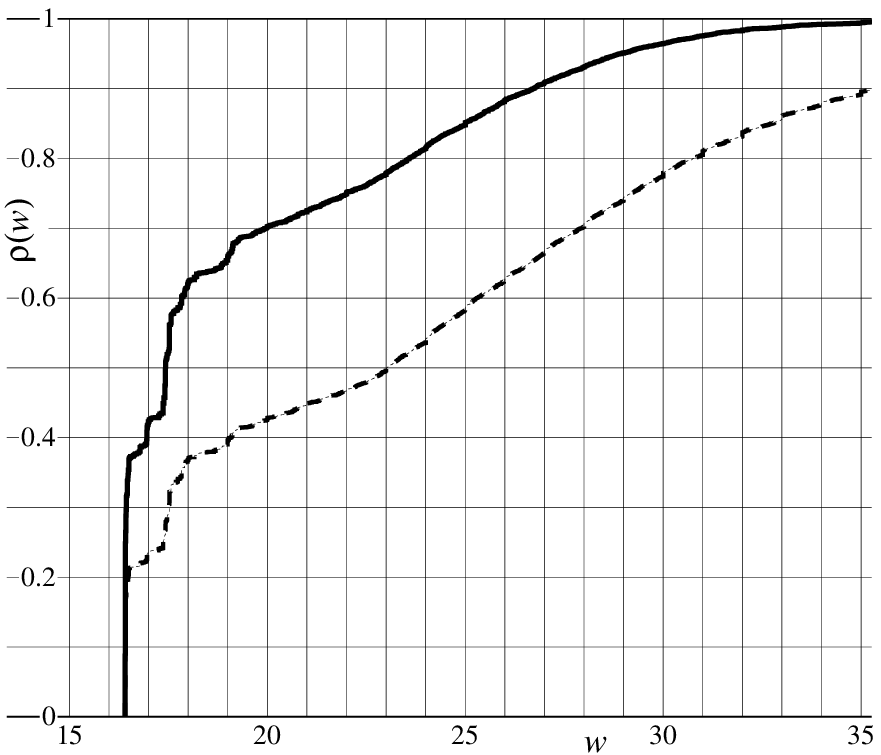}
 \caption{The probability/frequency of occurrence, $\rho(w)$, of the
pseudo-codewords with effective weight $w$ or smaller for the Tanner
$[155, 64, 20]$ code \cite{tannercode}. Solid and dashed lines represent
results of $10^4$ trials of MOA and PCS algorithm of \cite{08CS}
respectively.}
 \label{fig:spectra}
\end{figure}

We tested MOA on the popular example of the Tanner $[155, 64, 20]$
code \cite{tannercode}. The results are shown in Fig.~\ref{fig:spectra}.
We analyzed effective distance, $w$, of the pseudo-codewords found in
the result of $10^4$ trials (different in initial orientation). As in
the case of the PCS of \cite{08CS}, the probability (frequency),
$\rho(w)$, of finding pseudo-codeword with effective distance smaller
than $w$, grows monotonically with $w$. Like PCS, MOA result for the
smallest effective distance of the code is, $w_{\rm min} \approx 16.4037 
< 20$, where $20$ is the Hamming distance of the code. However, we also
observe that MOA is sampling the low-weight ``nearest-neighbor"
pseudo-codewords more efficiently than PCS, which is seen in
a steeper dependence of $\rho(w)$ as a function of $w$ in
Fig.~\ref{fig:spectra}. As discussed above, we attribute the better performance of MOA to
stronger bias towards the zero codewords convergence, as well as simpler and more
homogeneous (in the low weight sector of the pseudo-codewords)
initiation procedure.

\section{Conclusions and Path Forward}
\label{sec:Conclusions}

This paper reports new results related to analysis and algorithms
discovering the lowest-weight pseudo-codeword(s) of the LP decoding of
graphical codes performing over soft-output (log-concave) channels, like
the AWGN channel. On the theoretical side, we show here that the set of
correct decoding is a polytope in the space of noise. We also formulate
the problem of finding the smallest weight noise (instanton) as an
optimization problem, Eq.~(\ref{cone-opt}), looking for a maximum of a
convex function over a convex set (a polytope). The exact solution of
the problem is likely non-tractable, and we suggest heuristic iterative
algorithmic solution based on the majorization-minimization approach of the optimization theory \cite{04HL}.
We show that convergence of both MOA and PCS, introduced in \cite{08CS}, 
is monotonic. We also compare the
algorithms in simulations on the standard example of the Tanner $[155,
64, 20]$ code \cite{tannercode}, and observe that MOA is superior in discovering the low-weight part of the pseudo-codeword spectrum.

We plan to extend this research in the future along the following directions:
\begin{itemize}
\item Test MOA on other and longer codes.
\item Test MOA on other log-concave, but still binary, channels. We also envision extension of the technique to non-binary channels,  especially these related to phase modulation in modern fiber optics \cite{NKT2010}.
\item It will be useful to find a version of the majorization-minimization initiation which samples the ``nearest-neighbor" pseudo-codewords uniformly,  or (preferably) according to a given function of the effective weight.
\item The LP-decoding is a close relative of the generally faster but more difficult to analyze iterative BP-decodings. It will be useful to extend the polytope theory and the MOA algorithm discussed in the paper to the case of iterative decodings, for example to the basic min-sum algorithm.
\item Our major long-term goal consists in designing better graphical codes. We anticipate that MOA will be instrumental in searching over candidate codes (for example sampled from a properly expurgated ensemble of LDPC codes \cite{RiU05}) for the one showing the lowest error-floor possible.
\end{itemize}

\section{Acknowledgments}

The work of MC at LANL was carried out under the auspices of the
National Nuclear Security Administration of the U.S. Department of
Energy at Los Alamos National Laboratory under Contract No.
DE-AC52-06NA25396. MC acknowledges support of NMC via the NSF
collaborative grant CCF-0829945 on ``Harnessing Statistical Physics for
Computing and Communications.'' The work of MS is supported by NSF grant
DMS-0807592 ``Asymptotic Performance of Error Correcting Codes''.

\bibliographystyle{IEEEtran}
\bibliography{new_instanton,ref}

\end{document}